\documentclass[preprint,prb,preprintnumbers,amsmath,amssymb]{revtex4}
\usepackage{graphicx}% Include figure files
\usepackage{dcolumn}% Align table columns on decimal point
\usepackage{bm}% bold math
\newcommand{\Mat}[1]{{{\boldsymbol{#1}}}}

\def\be{\begin{equation}}
\def\ee{\end{equation}}
\def\dd{\mathrm{d}}
\begin{document}
\vspace{3cm}

\title{\bf Gravity as Archimedes' thrust and a bifurcation in that theory}

\author{Mayeul Arminjon}
%\email{arminjon@hmg.inpg.fr}
\affiliation{Laboratoire ``Sols, Solides, Structures'' \\
(CNRS / Universit\'e Joseph Fourier / Institut National Polytechnique de Grenoble) 
\\ BP 53, F-38041 Grenoble cedex 9, France. \\
Email: arminjon@hmg.inpg.fr\\ \newline \newline}

\begin{abstract}
 {\small Euler's interpretation of Newton's gravity (NG) as Archimedes' thrust in a fluid ether is presented in some detail. Then a semi-heuristic mechanism for gravity, close to Euler's, is recalled and compared with the latter. None of these two ``gravitational ethers" can obey classical mechanics. This is logical since the ether defines the very reference frame, in which mechanics is defined. This concept is used to build a scalar theory of gravity: NG corresponds to an incompressible ether, a compressible ether leads to gravitational waves. In the Lorentz-Poincar\'e version, special relativity is compatible with the ether, but, with the heterogeneous ether of gravity, it applies only locally. A correspondence between metrical effects of uniform motion and gravitation is assumed, yet in two possible versions (one is new). Dynamics is based on a (non-trivial) extension of Newton's second law. The observational status for the theory with the older version of the correspondence is summarized.}

\end{abstract}

\keywords{}
\maketitle
%

%\end{center}

% ----------------------------------------------------------------
%%%%%%%%%%%%%%%%%%%%%%%%%%%%%%%%%%%%%%%%
\section{Introduction}
%%%%%%%%%%%%%%%%%%%%%%%%%%%%%%%%%%%%%%%%
We now believe that Science can at best tell how things happen, not why. This is to say that phenomenologically efficient theories are preferred over theories that claim to have a ``real explanation" for physical phenomenons, but which are unable to give a quantitative account of experiments. I share this common opinion, of course. There cannot be any final explanation for any class of physical facts, simply because physical facts are precisely defined only inside some theoretical framework. Physics just builds theoretical models, each for some domain of physical reality, and which merely provide an approximate description of how things are going on. The better models are those that apply to larger domains of physical reality, and the ones that more closely approximate observations. (These are two conflicting constraints.)\\

Thus, Newton's theory of gravitation was obviously a phenomenological theory: using his mechanics, he derived it from Kepler's laws---themselves a kind of (very admirable) fitting of Tycho Brahe's precise observations. Moreover, Newton did not assume that the attraction at a distance, which he was postulating in his theory, did exist as such in the physical world: 

{\small``[I] use the words attraction, impulse, or propensity of any sort towards a centre, promiscuously, and indifferently: one for another; considering those forces not physically, but mathematically: wherefore the reader is not to imagine that by those words I anywhere take upon me to define the kind, or the manner of any action, the causes or the physical reason thereof, or that I attribute forces, in a true and physical sense, to certain centres, (which are only mathematical points); when at any time I happen to speak of centres as attracting, or as endued with attractive powers."} \cite{Newton1} 

In fact, he considered that his phenomenological attraction force might possibly result from the pressure of an ``aether" (I did read some careful sentence by him in this sense, but could not find it again), though he did not develop this idea. 
\footnote{\,
It is relevant to note that Newton was a close friend of Fatio de Duillier, the first proponent \cite{Lunteren} of what became known as Le Sage's theory and that is now experiencing a revival of interest under the name ``pushing gravity theory." \cite{Edwards} According to this concept, gravity would be due to the massive bodies {\it shielding} an ubiquitous flux of ``ultramondane particles." As far as I know, an accurate model of gravity based on Fatio--Le Sage's concept is still lacking.
}
\\

But {\it Euler} did imagine a definite mechanism for gravity and did describe it at length, \cite{Euler1, Euler2} although that work of Euler is not well-known. Now the aim of the present paper is to discuss in more detail the tentative mechanism for gravity, on which is based a formerly developed theory of gravity---\cite{A8,A20} and also, to somewhat extend that theory. Since that mechanism turns out to be quite close to that imagined by Euler (although I was not aware of that for my first work on this theory \cite{A8,A9}), I shall first review his relevant work in some detail (Sect. \ref{Euler_pressure}). In Sect. \ref{MA_pressure}, I shall present my own version of a semi-heuristic mechanism for gravity and shall compare it with Euler's concept. Section \ref{MA_equivalence} will expose the idea of a correspondence between metrical effects of motion and gravitation, as well as the two possibilities arising therefrom as regards the space-time metric. The possible forms of the scalar field equation shall be discussed in Sect. \ref{FieldEquation}. Then the dynamics of the theory will be summarized (Sect. \ref{Dynamics}). The current state of the experimental test shall be reviewed in Sect. \ref{Experiment}, and my conclusion makes Sect. \ref{Conclusion}.\\

%%%%%%%%%%%%%%%%%%%%%%%%%%%%%%%%%%%%%%%%%%%%%%%%%%%%%%%%%%%%%%%%%%%%%%%%%%%
\section{Euler's interpretation of Newton's theory} \label{Euler_pressure}
%%%%%%%%%%%%%%%%%%%%%%%%%%%%%%%%%%%%%%%%%%%%%%%%%%%%%%%%%%%%%%%%%%%%%%%%%%%

In a first paper, \cite{Euler1} Euler admitted that the ``{\small least parts of matter}" or ``{\small molecules}" have all the same density---so that the different densities of material bodies are due to the fact that the bodies are dominantly composed of ``{\small pores}"--- and that these ``{\small molecules}" are surrounded by ``{\small an extremely subtle matter which, by its motion, is endowed with a force capable of pushing the bodies downwards and of producing all phenomenons of gravity.}" (My translation.) The ``motion" referred to must actually be thought of as closely related to the {\it pressure} exerted by the ``{\small subtle matter}": 

``{\small Now, in whatever way we imagine the cause of gravity, as it is the effect of the pressure of a fluid, the force with which each molecule is pushed will always be proportional to the extension or the volume of that molecule. Indeed it is a general rule of hydrostatics that fluids act according to the volumes: a body immersed in water is always pushed by a force equal to the weight of an equal volume of water, but in an opposite direction.}" 

Thus, in effect he a priori admitted that gravity is due to Archimedes' thrust exerted by a ``{\small subtle fluid}," and, in order that this force reduce to a mass force, as does gravity, he had to admit also that the finest components of any matter all have the same density. The main part of the usual, weighty bodies had then to be made of ``{\small pores}," which had to be occupied by the assumed fluid, and Euler asked whether that fluid which, ``{\small however subtle it be, will yet be material}," was of the same kind as the weighty matter. But if that would be the case, 

``{\small the whole space would be filled with a matter everywhere equally dense, and even denser as gold; which would make very difficult, not to say impossible, the explanation of motion. Indeed, although there is only a small part among the bodies which are weighty bodies and which can be perceived by phenomenons, the other part, owing to its very large density, could but resist to motion; now we do hardly notice any resistance by which the motion of bodies be diminished, as soon as we have removed the resistance of weighty bodies, like the air.}"

For that reason, he sustained that

``{\small the matter which constitutes the subtle fluid, cause of the gravity, is of an utterly different nature from the matter, of which all sensible bodies are composed. There will hence be two kinds of matter, one which provides the stuff to all sensible bodies, and of which all particles have the same} [high] {\small density [...]; the other kind of matter will be that of which the subtle fluid, which causes gravity, and which we name ether, is composed of. It is probable that this matter has always the same degree of density, but that this degree is incomparably smaller than that of the first kind.}"\\

In the second paper, \cite{Euler2} Euler explained in more detail the mechanism by which the pressure of the ether or subtle fluid would cause gravity:

``{\small Those who attribute gravity to an attractive force of the Earth base their opinion mainly on the fact that otherwise no origin could be displayed for this force. But since we proved that all bodies are surrounded with ether and are pressed by the elastic force of the latter, we do not need to search elsewhere the origin of gravity. Only if the pressure of the ether would be everywhere the same, which assignment is indistinguishable from that of its equilibrium, would the bodies be equally pressed from every side, and thus would not be induced in any motion. But if we assume that the ether around the Earth is not in equilibrium, and that instead its pressure becomes smaller as one comes closer to the Earth, then any given body must experience a stronger pressure downwards on its superior surface that it does upwards on its inferior surface; it follows that the downwards pressure will have the advantage and hence that the body will really be pushed downwards, which effect we call {\it gravity}, and the downwards-pushing force the {\it weight} of the body.}" (My translation.) 

Thus, gravity would indeed be Archimedes' thrust due to the gradient of the ether pressure, which would act on the ``{\small basic particles}" (``{\small grobe Theilchen}") [those that ``{\small provide the stuff to all sensible bodies}," see the paragraph before]. This force depends only on the volume. More precisely, Euler {\it assumed} that, for a point being at a distance $x$ from the center of the Earth, the pressure is 
\be \label{pressureloss}
p_e=h-\frac{A}{x},
\ee
where $h$ is ``{\small the height, by which we let the pressure of the ether be expressed, when it is at rest}" [i.e., in equilibrium]. He deduced therefrom that the weight of a body is 
\be \label{Eulerweight}
P = \frac{A}{x^2}c^3, 
\ee
with $c^3$ its ``true volume" (``{\small wahre Gr\"osse}"), i.e., the sum of the volumes of the ``{\small basic particles}" in it. Thus, he ``deduced" Newton's inverse square law from the spatial variation (\ref{pressureloss}) of the pressure---or rather the opposite way. He then noted that this expression of the weight means that the corresponding pressure ``loss" $-\frac{A}{x}$ is very large: he expressed the pressure by the height of a column of water (thus 32 feet for the ``{\small elastic force of the air}," i.e., the atmospheric pressure), and he expressed the weight by that of a volume of water on the Earth's surface. In such units, the weight on the Earth's surface is $c^3 \rho_p$ with $\rho_p$ the common mass density of the ``{\small basic particles},"
\footnote{\,
Except for $p_e$ and $\rho_p$, which I also use in my own work, and for $M_1$, $M_2$ below, the notations in this Section are those used by Euler himself.
}
hence we have from (\ref{Eulerweight}):
\be \label{rho_p}
\rho_p = \frac{A}{r^2}
\ee
with $r$ the Earth's radius. Since $\rho_p>19$ (the density of gold), he thus found that $A > 19 r^2$. On the Earth's surface, the pressure ``lost" (as compared with the pressure of the ether in equilibrium at infinity, given by the height $h$) is hence, by (\ref{pressureloss}), $-\delta p_e = \frac{A}{r} > 19 r$, which is indeed more than $10^7$ atmospheres.\\

When there are several celestial bodies, as is indeed the case in reality, Euler remarked that the pressure losses caused by each of them should add up, so that the pressure is in fact 
\be \label{pressureloss2}
p_e=h-\frac{A}{z} -\frac{B}{y}-\frac{C}{x}-\frac{D}{v} -\mathrm{etc.},
\ee
instead of (\ref{pressureloss}), where $z$, $y$, $x$, $v$, etc., are the distances to the respective centres of the celestial bodies. This, of course, is equivalent to assuming that the weight (the gravitational force) is the sum of the attraction forces caused by the different bodies. Euler noted that $A$, $B$, etc., should have the form $A=mM_1$, $B=mM_2$, etc., where $M_1$, $M_2$, etc., are the masses of the bodies, and the constant $m$ may be calculated from (\ref{rho_p}) (assuming one knows $\rho_p$, in fact he assumed $\rho_p =40$ for illustrative purpose). He concluded:

``{\small But although we must stay here and can hardly hope to ever be able to elucidate the true origin of the decrease in the elastic force of the ether, one can yet more easily accept this, rather than to barely admit that all bodies are endowed by Nature with an ability to attract one another. Indeed one cannot get the slightest beginning of an understandable concept of this attraction, whereas, on the contrary, one can to the very least acknowledge it as possible that the elastic force of a fluid be diminished, and one conceives also that this might happen by virtue of some law of Nature. But all rests on the following two points: firstly, why is the pressure of the ether decreased by the presence of the basic bodies which it contains? And secondly, why does this decrease become larger and larger, as one comes closer to the body? The reason for this must thus apparently be in the basic matter} [i.e. the union of the ``basic particles"]{\small, of which any body is made, and the basic matter must cause a motion in the ether, by which the equilibrium is raised. If one has first got so far, then it is easy to show that the pressure of the ether should be reduced in proportion.}"

%%%%%%%%%%%%%%%%%%%%%%%%%%%%%%%%%%%%%%%%%%%%%%%%%%%%%%%%%%%%%%%%%%%
\section{A neo-Eulerian mechanism for gravity} \label{MA_pressure}     
%%%%%%%%%%%%%%%%%%%%%%%%%%%%%%%%%%%%%%%%%%%%%%%%%%%%%%%%%%%%%%%%%%%

The starting point for this work was Romani's concept of a perfectly-fluid ``constitutive" ether, according to which the elementary constituents of matter should be nothing else than organized flows in the ether, like vortices. \cite{Romani} This concept allows one to get a picture of the creation/ annihilation/ transmutation of particles, and of the numerous instable ``resonances," all observed in particle physics---but, of course, it would be very ambitious to attempt a reconstruction of particle physics along this line. Romani also considered that gravity is due to a gradient in the ether density \cite{Romani} (Vol. 1). He stated that the ether density should increase towards the Sun and cause a light deflection in the way it occurs in an optical medium with variable index, thus following Fermat's principle \cite{Romani} (Vol. 2). This same idea has been proposed by several other authors, e.g. Podlaha \& Sj\"odin. \cite{PodlahaSjodin} However, as such it does not provide any interpretation of Newton's attraction, nor indeed any mechanism for usual gravity. I tried to find one.

\subsection{Gravity Acceleration vs. Ether Pressure} \label{Archimedes}

The successful attempt used the simplest concept: only a perfect fluid could fill the space without braking the motion of material bodies, and only the pressure force can be exerted by a such fluid. \cite{A8} I had first rejected that concept due to its a priori shocking consequence, namely the fact that the pressure $p_e$ of the ether has to {\it decrease} towards the attracting centre. The resultant of the pressure forces on a small object $\omega$ is Archimedes' thrust $\mathbf{F}_\mathrm{A}=-V(\omega)\mathrm{grad}p_e$, where $V(\omega)$ is the {\it volume} of that object and $\mathrm{grad}p_e$ is the local value of the pressure gradient. In order that this force depend in fact on the {\it mass} of the object, one thus finds that the elementary particles, which make matter, and which are the objects that are actually subjected to gravity, must all have the same density $\rho_p$. The value of the gravity acceleration is thus \cite{A8}:
\be \label{g_rho_p}
\mathbf{g}=-\frac{\mathrm{grad}p_e}{\rho_p},
\ee
which turns out to be equivalent to Euler's assumption \cite{Euler1} if $\rho_p$ is assumed constant (see Sect. \ref{Euler_pressure}). At that point, however, I noted that $\rho_p$ may actually depend on the ether pressure $p_e$, but only on $p_e$, and that it would seem miraculous, unless the particles themselves are made of ether; which I did assume, thus staying with Romani's constitutive ether, and hence setting
\be \label{g_rho_e}
\mathbf{g}=-\frac{\mathrm{grad}p_e}{\rho_e},
\ee
where $\rho_e = \rho_e(p_e)$ is the ``density" in the ether, the latter being assumed a barotropic fluid. Hence, unless that fluid is incompressible, i.e. unless $\rho_e = \mathrm{Const.}$ (which is a special, degenerate case of a barotropic fluid), Eqs. (\ref{g_rho_p}) and (\ref{g_rho_e}) are not equivalent.

\subsection{The Ether of Gravity Does not Obey classical Mechanics}

The concept of the constitutive ether means that there would be nothing but the perfectly-fluid ether, hence in that case the ether should not ``brake" the motion of elementary particles. In fact it seems that it would be also the case if the perfectly-fluid ether would only surround elementary particles which would be made of a different stuff (as assumed by Euler), and this independently of the high or low density of the fluid: indeed a truly perfect fluid does not brake objects; this result is known in classical fluid mechanics as ``d'Alembert's paradox," although it is no paradox (Ref. \cite{L&L_fluides}, Sect. 11). But, of course, according to our present understanding of particle physics, classical mechanics does not apply at the scale of elementary particles. It is interesting that the model itself will give us a warning about applying classical mechanics. First, it is easy to see that the ether assumed in either mechanism is not in equilibrium. Let us first consider the space-filling constitutive ether assumed by me, that includes the elementary particles of matter considered as flows in the ether. It is subjected only to the pressure force, which is equivalent to the internal force $-\mathrm{grad}p_e$ per unit volume. Considering now Euler's ether, which is distinct from the ``{\small basic particles}," the same is true, but the unit volume is that of ether, not the total volume (ether plus basic particles). Therefore, if either ether were to obey classical fluid mechanics, it would have to be in a {\it motion}, with velocity $\mathbf{u}_e$, obeying... Euler's equation, plus the ``mass" conservation. In the case of Euler's ether, these equations are not immediate to write at the macroscopic scale, for we have then a two-phase medium. (It is the macroscopic scale which is relevant here, because otherwise one has different equations, depending on which phase is involved, and because gravitation is a macroscopic force: see the following Subsection.) But let us first discuss the case with the constitutive ether. Euler's equation is then simply
\be \label{Euler_e}
\rho_e \frac{\dd \mathbf{u}_e}{\dd T} \equiv \rho _e \left(\frac{\partial \mathbf{u}_e}{\partial T}+ (\mathbf{grad\,u}_e)\mathbf{.u}_e \right) = -\mathrm{grad}p_e, 
\ee
and the conservation of that fluid should then be written as the usual continuity equation:
\be \label{continuity}
\partial_T \rho_e + \mathrm{div}(\rho_e \mathbf{u}_e) = 0.
\ee
But the ether density $\rho_e$ is directly related to the gravity acceleration $\mathbf{g}$ by Eq. (\ref{g_rho_e}), and $\mathbf{g}$ must in fact be determined by the positions of the massive bodies. For instance, let us demand that Newton's gravity be {\it exactly} valid (as Euler demanded). Then we introduce the Newtonian potential $U$, governed by Poisson's equation:
\be \label{Poisson}
\Delta U = -4\pi G \rho,
\ee
with $\rho$ the {\it macroscopic} density of matter and $G$ the gravitational constant. And, assuming that macroscopic matter, like the ether, is in the form of a barotropic perfect fluid (for simplicity), it has a velocity $\mathbf{u}$, a pressure $p$, with $\rho = \rho(p)$, which fields obey also Eqs. (\ref{Euler_e}) and (\ref{continuity}), though without the index $e$, and with the gravitational force on the r.h.s. of the former, that is
\be \label{FP}
\rho \frac{\dd \mathbf{u}}{\dd T} = - \mathrm{grad}p +\rho\,\mathrm{grad}U, \quad \partial_T \rho + \mathrm{div}(\rho \mathbf{u}) = 0.
\ee
Thus we have {\it nine} independent scalar unknowns, namely $U,p,\mathbf{u},p_e,\mathbf{u}_e$. Now, since $\mathbf{g}=\mathrm{grad}U$, (\ref{g_rho_e}) is equivalent to the scalar equation
\be \label{U_p_e}
U = -G(p_e)+\varphi(T), \quad  G(p_e) \equiv {\small \int} dp_e/\rho_e(p_e).
\ee
But even then we have still {\it ten} independent scalar equations, namely (\ref{Euler_e}), (\ref{continuity}), (\ref{Poisson}), (\ref{FP}), and (\ref{U_p_e}). We can get from (\ref{g_rho_e}) (or from (\ref{U_p_e}))
and from (\ref{Euler_e}):
\be \label{accel_e}
\frac{\dd \mathbf{u}_e}{\dd T} = \mathrm{grad}U, 
\ee
which means that the ether would have to be in a free fall in the gravity acceleration field. Although the equation of motion replacing (\ref{Euler_e}) and the ``mass" conservation replacing (\ref{continuity})
are more complicated to write (at the macroscopic scale) for Euler's ether, they must also add four scalar equations. Hence, the overdetermination is also true for Euler's ether, for which one uses Eq. (\ref{g_rho_p}) instead of (\ref{g_rho_e}). The overdetermination of the field equations means that usually they will have no solution. I conclude that the ether, in which gravity is Archimedes' thrust, can hardly obey classical mechanics.

\subsection{The Micro-Ether and the Macro-Ether} \label{micro-macro}

In fact I do not find it surprising that classical mechanics does not apply to the gravitational ether. If it would apply, one would have to answer the question: but how are the inertial frames defined, if the ether has a general fluid motion? It seems indeed that the introduction of an ether should enable us to define an inertial frame---thus explaining the origin of the inertial frames, which is a mysterious question in Newton's mechanics. Moreover, gravity is a macroscopic force, in the sense that the gravity acceleration vector of Newton's gravity is known to vary significantly only over macroscopic distances, due to the variation of the position with respect to big, massive bodies---in sharp contrast with the other known forces (electromagnetic and nuclear forces), which vary much more rapidly with distance. (This does not contradict the fact that gravitation is felt even by the finest particles, of course.)\\

Therefore, the ether of gravity must be the macroscopic ether or ``macro-ether," obtained by averaging the microscopic fields (velocity $\mathbf{u'}_e$, pressure $p'_e$, density $\rho'_e$) that characterize the assumed perfectly-fluid ether, which I call the ``micro-ether." In particular, we may define a reference frame E as the one whose each point has as its velocity the local value of the averaged velocity field $\mathbf{u}_e(T,\mathbf{x}) \equiv \langle\langle\mathbf{u'}_e\rangle\rangle_{T,\mathbf{x}}$ (the velocity field of the micro-ether, $\mathbf{u'}_e$, being taken in any reference frame, possibly a deformable or ``fluid" reference frame, see Ref. \cite{A8} for a detailed study of that notion). Then, by definition of the frame E, the average velocity field $\mathbf{u}_e$ is identically {\it zero} in that very frame. It is that frame which will define the equivalent of Newton's absolute space. Thus, because the macro-ether defines the reference frame in which mechanics is primarily defined, the motion of that reference frame itself does not obey mechanics. And gravitation is assumed to result from the {\it macroscopic} part of the pressure gradient in the micro-ether: the microscopic pressure and velocity fields $p'_e$ and $\mathbf{u'}_e$ would then have to account for microphysics, indeed---but this is another story, which I do not try to tell (one may tentatively assume that the microscopic motion of the ether with respect to its mean rest frame E does obey mechanics \cite{A8}).\\

In order to build a {\it self-consistent} theory of gravity based on the foregoing considerations, I consider henceforth the macro-ether or preferred reference frame as a primary concept, 
\footnote{\,
Hence, it is not needed to define precisely the average fields $\langle p'_e\rangle_{T,\mathbf{x}}$ and $\langle\mathbf{u'}_e\rangle_{T,\mathbf{x}}$. Tentatively, one may define $p_e(T,\mathbf{x}) \equiv \langle p'_e\rangle_{T,\mathbf{x}}$  as the volume average in a finite space domain around $\mathbf{x}$, and define $\langle\langle\mathbf{u'}_e\rangle\rangle_{T,\mathbf{x}}$ by the same volume average, followed by a time average in a finite time interval around $T$ (Ref. \cite{A28}, Subsect. 2.2).
}
and I assume that the gravity acceleration is defined by Eq. (\ref{g_rho_e}). The field $p_e$ has to obey some partial differential equation which shall be chosen mainly by phenomenological means, i.e., from the wish to account in a mathematically simple way for what we know about effective gravity (or rather for what we think we know). But since we start from an equation for the gravity acceleration, we shall have to define and use a dynamics based on some extension of Newton's second law of motion, rather than on Einstein's assumption according to which ``free test particles follow space-time geodesics."

\subsection{A Constraint on the Equation for the Field $p_e$} \label{constraint}

The most obvious constraint to be imposed on the equation for the field $p_e$ is that Newton's gravity should be recovered in some limit, the question is: in which limit? Since we assume that gravitation results from the macroscopic part of the pressure gradient in the micro-ether, we expect that, if that fluid is (macroscopically) compressible, a disturbance in the (macroscopic) ether pressure should propagate with the ``sound" velocity,
\be \label{c_e}
c_e = \left(\frac{\dd p_e}{\dd\rho_e}\right)^{1/2}. 
\ee              
On the other hand, we know that Newton's gravitation propagates instantaneously. Therefore, it should correspond to the limiting case of an incompressible fluid. The latter should be an excellent approximation in many cases, because Newton's gravity is extremely accurate in many situations (in particular in the solar system). Newton's gravity is characterized by Poisson's equation for the gravity acceleration $\mathbf{g}$:
\be \label{Poisson_g}
\mathrm{div}\, \mathbf{g} = -4 \pi G \rho.                        
\ee
Together with Eq. (\ref{g_rho_e}) for $\mathbf{g}$, the requirement that Eq. (\ref{Poisson_g}) is recovered in the incompressible case, i.e., $\rho_e = \mathrm{Const.}$, leads immediately to an equation for the field $p_e$:
\be \label{p_e_Newton}
\Delta p_e = 4 \pi G \rho \rho_e,                     
\ee
which should thus apply in the limit of the degenerate barotropic relationship $\rho_e(p_e)= \mathrm{Const}$.

%%%%%%%%%%%%%%%%%%%%%%%%%%%%%%%%%%%%%%%%%%%%%%%%%%%%%%%%%%%%%%%%%%%
\section{Correspondence between metrical effects of motion and gravitation: the bifurcation}\label{MA_equivalence}     
%%%%%%%%%%%%%%%%%%%%%%%%%%%%%%%%%%%%%%%%%%%%%%%%%%%%%%%%%%%%%%%%%%%

Surely, nowadays a theory of gravitation should account in some way for special relativity (SR), because SR has made a lot of experimentally-confirmed predictions, the set of which is very hard to obtain without SR. This is well-known. The minimum requirement for a theory of gravity to account for SR is that SR should be exactly recovered when gravity evanesces, i.e., in the limit $G \rightarrow 0$. \cite{Will}

\subsection{Special Relativity with an Ether}

It is a historical curiosity that the theory which we now name SR, on one hand has been entirely derived, in one version, by physicists (mainly Lorentz and Poincar\'e) who did accept and explicitly used Lorentz's concept of the ether as a preferred inertial frame in which Maxwell's equations are valid, and on the other hand, in its best-known version, is still currently believed to have discarded that very concept! Indeed Einstein wrote in his celebrated paper, \cite{Einstein1905} which initiated the Einstein-Minkowski version of SR, that ``the introduction of a `luminiferous ether' will appear superfluous." But everyone interested can and should check that the ``postulate of relativity" and all basic features of SR (including the Lorentz transformations for positions and for velocities; the Lorentz invariance of the Maxwell equations; relativistic dynamics with velocity-dependent mass; relativistic 4-vectors such as energy-momentum, charge-current, electromagnetic potential; the Lorentz group and its invariants, among them the quadratic form $x^2+y^2+z^2-t^2$; and the 4-dimensional space-time with coordinates $x,y,z,t\sqrt{-1}$) are written in Poincar\'e's papers. \cite{Poincare1905,Poincare1906} (Both papers are available online.)
\footnote{\,
It is easier yet to check this in the commented translation by Logunov. \cite{Logunov95}
}
 It can also be easily checked that Poincar\'e, and Lorentz as well, considered an ether in their papers that founded the Lorentz-Poincar\'e version of SR, \cite{Poincare1906,Lorentz} and persisted until they died in 1912 and 1928 respectively---although, before 1900, Poincar\'e had been close to abandon the ether; thus, he considered abandoning the concept of an ether, but came back to it precisely in the context of SR. His reason may have been that, without an ether, there is no preferred time any more, so that simultaneity becomes inherently and incurably a relative notion. \\

Since one version of SR has been derived by persistent ether theorists of the stature of Lorentz and Poincar\'e, it is a priori obvious that SR should be fully compatible with Lorentz's concept of ether. That this is indeed the case, has been proved with a luxury of details by Prokhovnik, \cite{Prokhovnik67, Prokhovnik85} J\'anossy, \cite{Janossy} Pierseaux, \cite{Pierseaux} and Brandes. \cite{Brandes01} The main difference between the Lorentz-Poincar\'e and the Einstein-Minkowski versions of SR is this: in the latter version, the effects on distances and time intervals, associated with the Lorentz transformation of space and time coordinates, are seen as a mere effect of perspective in space-time. Whereas, in the former version, they are seen as true effects which result from a real contraction of physical objects that move through the fundamental inertial frame (Lorentz's ether). That contraction implies the slowing down of a moving ``light clock" (an interferometer's arm). Since the Michelson-Morley experiment means that such light clocks do measure physical time, it follows that the time flows more slowly in a moving frame. See Ref. \cite{Prokhovnik67} for a rather complete construction of SR along this line, and cf. Ref. \cite{A9} for an outline. Because, in the Lorentz-Poincar\'e version, SR follows from this ``true" Lorentz contraction which is a dynamical effect, SR does not necessarily apply to the whole of physics. More exactly, what is needed in that derivation of SR (as well as in the derivation based on Einstein's two well-known assumptions) is that {\it physical} space and time are {\it homogeneous}, i.e., the physical clocks and the measuring rods behave in the same way at any time and at any place. Now, the interpretation (\ref{g_rho_e}) of gravity as a {\it gradient} of pressure and density in the {\it universal fluid} naturally seems to imply that gravitation makes physical space and time {\it heterogeneous}.
\footnote{\,
Note that this is also the case in general relativity (GR), according to which SR holds true only in infinitesimal domains of space and time. 
}
Therefore, we expect that SR and its exact Lorentz invariance should be broken by the presence of a gravitational field. Precisely, the theory of gravity \cite{A8}--\cite{A20} is a preferred-frame theory, and most of its equations are valid only in an a priori postulated preferred reference frame or ``macro-ether" (see Subsect. \ref{micro-macro} for the explanation of that name). Now a bifurcation allowing some extension of that theory will be recognized.

\subsection{Two alternative Assumptions about metrical Effects of Gravity} \label{correspondence}

When combined with the Lorentz-Poincar\'e version of SR, the assumption of a heterogeneous gravitational ether leads naturally to assume that a gravitational field, i.e. a heterogeneous field of ether pressure $p_e$ and density $\rho_e=\rho_e(p_e)$, does indeed modify the behaviour of clocks and rods. To see this, let us imagine that an observer, say Nancy, is moving through a homogeneous ether, thus in the absence of gravity. Her measuring meters are Lorentz-contracted, parallel to the direction of her velocity $\mathbf{u}$ with respect to the ether, so that a given domain in the ether has, for her, a greater volume $\delta V'= \gamma _u \delta V$, where $\delta V$ is the volume of the domain, when this domain is seen from the ether frame, and where $\gamma _u>1$ is the Lorentz factor.
\footnote{\,
This is true if one uses the ``absolute" simultaneity, i.e., that obtained by Poincar\'e-Einstein clock synchronization in the ether frame; cf. Ref. \cite{A18}. As indicated by the work of Mansouri \& Sexl, \cite{MansouriSexl} and as shown in detail by Selleri, \cite{SelleriFP,Selleri03} a theory observationally very close to SR can be based on using just that notion of simultaneity. Selleri's theory is yet different from the Lorentz-Poincar\'e version of SR, which uses the standard (Poincar\'e-Einstein) synchronization convention, and which is mathematically and observationally equivalent to Einstein-Minkowski SR---except for the fact that Lorentz-Poincar\'e SR, as Selleri's theory, and in contrast with Einstein-Minkowski SR, would not be {\it falsified} by the observation of genuine superluminal signal velocities: since, in Lorentz-Poincar\'e SR, the Poincar\'e-Einstein conventional simultaneity is regarded as ``true" only in one reference frame (the ether), it follows that Lorentz-Poincar\'e SR is in advance compatible with the ``apparent" inversions of temporal successions in different frames, that would arise in the case of superluminal signal velocities. However, in such case, Selleri's theory would have the advantage over Lorentz-Poincar\'e SR that it would {\it describe} the observed absolute simultaneity.}
Therefore, Nancy would be led to find that the ether density is smaller for her than for a fixed observer, $\rho'_e=\rho_e/\gamma_u$. Thus, the Lorentz contraction of her meters and the Larmor dilation of the period of her clock depend on precisely the ratio of the ether density in the moving frame to that in the ether frame. \\

It is hence natural to assume that, in a heterogeneous ether, thus with a gravitational field, the clocks will be slowed down and the meters will be contracted, in precisely the ratio of the local ether density $\rho_e$ to the density $\rho_e^\infty$ in the region far enough from massive bodies and thus free from gravitational field,
\be \label{beta}
\beta(\mathbf{x},T) \equiv \rho_e(\mathbf{x},T) /\rho_e^\infty(T).
\ee
($\rho_e^\infty$ may be precisely defined as $\rho_e^\infty(T) \equiv \mathrm{Sup}_{\mathbf{x}\in\mathrm{[space]}}\rho_e(\mathbf{x},T)$, hence it generally depends on the time $T$.) The very notion of a clock slowing and a contraction of objects implies that there must be a {\it reference} with respect to which these effects do occur, and the simplest possibility is clearly to assume a flat reference metric. Thus, space-time is endowed with two metrics: a flat ``background" metric $\Mat{\gamma}^0$ and a curved ``physical" metric $\Mat{\gamma}$. More precisely, I assume that there are Galilean coordinates $(x^\mu )$ for $\Mat{\gamma}^0$ [i.e., in that coordinates, $(\gamma^0_{\mu \nu }) = \mathrm{diag}(1, -1, -1, -1)$], which are adapted to the preferred frame E. (``Adapted coordinates" are such that any particle bound to the given frame has constant space coordinates.) The inertial time in the preferred frame, $T \equiv x^0/c$, is called the ``absolute time." Due to the assumed time-dilation, the expression of the physical metric in any coordinates $(y^\mu)$ adapted to the frame E, with $y^0 = x^0 = cT$, is:

\begin{equation} \label {spacetimemetric}
\dd s^2 = \gamma_{\mu\nu} \dd y^\mu \dd y^\nu = \beta^2 (\dd y^0)^2 - g_{ij} \dd y^i \dd y^j,
\end{equation}
where $\Mat{g}$ is the physical space metric in the frame E. Equivalently, we have $\gamma_{00} = \beta^2,   \gamma_{ij} = - g_{ij},    \gamma_{0i} = 0$ in such coordinates. \\

{\it But} there are two possibilities as regards the gravitational contraction: {\it either} we may assume \cite{A9} that, in the same way as for the case with a uniform motion, it occurs only in one direction---which then can only be the direction of the gravity acceleration $\mathbf{g}$, i.e., that of the density and pressure gradients, Eq. (\ref{g_rho_e}). In that case, the ``physical" space metric $\Mat{g}$ is found \cite{A19} to have the following relation to the Euclidean metric $\Mat{g}^0$ (which is the spatial part, in the frame E, of the flat space-time metric $\Mat{\gamma}^0$, as $\Mat{g}$ is for metric $\Mat{\gamma}$):
\begin{equation} \label {spacemetric}
		 			\Mat{g} =	\Mat{g}^0 +\left(\frac{1}{\beta^2}-1\right)\Mat{h}, \quad\Mat{h}\equiv\frac{\nabla \beta \otimes \nabla \beta}{(\nabla \beta)^2}      
\end{equation}
(with $(\nabla \beta)_i  \equiv \beta _{,i}$ and $(\nabla \beta)^2 \equiv g^{0ij} \beta _{,i} \beta _{,j}$, where $(g^{0ij})$ is the inverse matrix of the matrix $(g^0_{ij})$ of $\Mat{g}^0$ in the  coordinates $(y^i)$). {\it Or}, considering that space must remain isotropic, we may assume instead that the gravitational rod contraction in the ratio $\beta$ is the same in all directions (as Podlaha \& Sj\"odin \cite{PodlahaSjodin} apparently assumed). In that case, the relation between the flat and physical space metrics in the frame E will be simply
\begin{equation} \label {spacemetric2}
		 			\Mat{g} =	\beta^{-2}\Mat{g}^0.
\end{equation}

One may also imagine intermediate solutions, of course, but these are the two extreme and most natural possibilities. Until now, I have considered only the first (anisotropic) possibility (\ref{spacemetric}), for which the correspondence between metrical effects of motion and gravitation is the closest. But it turns out that it leads to a disturbing violation of the weak equivalence principle for an extended body at the point particle limit. \cite{A33,B23} As I show, \cite{A33} this violation comes from the fact that the ``anisotropic" spatial metric (\ref{spacemetric}), or rather the post-Newtonian (PN) approximation to this metric, contains terms that depend on the spatial derivatives of the Newtonian potential $U$; hence, that part of the PN acceleration which involves the spatial Christoffel symbols contains {\it second} spatial derivatives $U_{,i,j}$ of $U$---and the ``self" part of the $U_{,i,j}$'s (i.e., the contribution to $U_{,i,j}$ that comes from the body itself, whose the acceleration is being computed) does not evanesce with the size of the body. For that reason, I expect that the same violation should also occur in GR, depending on the gauge, \cite{A33,O2} but it would be more difficult to prove it, due to the complexity of GR (which makes it difficult \cite{O1} to develop, for PN calculations, asymptotic schemes in the line of those which are developed in applied mathematics). Coming back to the scalar ether theory, if one substitutes the ``isotropic" spatial metric (\ref{spacemetric2}) for the ``anisotropic" metric (\ref{spacemetric}), the PN approximation to (\ref{spacemetric2}) will involve $U$ but not the derivatives $U_{,i}$, hence the violation should not occur any more.

%%%%%%%%%%%%%%%%%%%%%%%%%%%%%%%%%%%%%%%%%%%%%%%%%%%%%%%%%%%%%%%%%%%
\section{Possible forms of the scalar field equation} \label{FieldEquation}
%%%%%%%%%%%%%%%%%%%%%%%%%%%%%%%%%%%%%%%%%%%%%%%%%%%%%%%%%%%%%%%%%%%

According to the correspondence postulated in Subsect. \ref{correspondence}, the metrical effects of a gravitational field depend on the ratio $\beta$, Eq. (\ref{beta}). Thus, the constraint that Newton's gravity should be recovered in the limit of an incompressible ether, i.e. for $\rho_e = \rho_e^\infty$, means that Eq. (\ref{p_e_Newton}) has to be recovered for $\beta =1$. In the case with a compressibility, we expect, as mentioned in Subsect. \ref{constraint}, that in general there should be pressure waves propagating with the velocity (\ref{c_e}). Hence, the left-hand side of the sought equation should then be a kind of wave operator involving the (a priori variable) propagation velocity $c_e$, instead of the Laplace operator---which should be recovered in the case of a static field, for which the propagation does not play any role. But the idea according to which material particles are microscopic flows in the ether implies that this ``sound" velocity should be a limit for the motion of material particles, and since SR sets the other limit $c$, one must have $c_e=c$ everywhere and at every time, which means that $p_e=c^2\rho_e$. It is, of course, when evaluated with the {\it physical} metric that the speed limit is $c$, hence the foregoing discussion applies to the equation for the scalar field, in terms of the physical metric; in particular, the Laplace operator is in terms of the physical, Riemannian space metric $\Mat{g}$. Thus, we are led to postulate an equation of the form
\be \label{possibleField}
\Delta_{\Mat{g}} p_e + (\mathrm{time\ derivatives\ of}\ p_e) =  4 \pi G \sigma \rho_e F(\beta), 
\ee
\be \nonumber
F(\beta) \rightarrow  1  \mathrm{\ as\ }  \beta \rightarrow 1,                   
\ee
the time-derivative term being such that, in the appropriate (``post-Minkowskian") limit, involving the condition $\beta \rightarrow 1$, the operator on the l.h.s. becomes equivalent to the usual (d'Alembert) wave operator. Moreover, the Newtonian density $\rho$ has been replaced by some mass-energy density $\sigma$, defined in terms of the energy-momentum tensor $\mathbf{T}$. This does not fix the equation, of course. Until now, and thus with the ``anisotropic" space metric (\ref{spacemetric}), I have postulated the following equation:
\begin{equation} \label{champ1}
\Delta_\Mat{g} p_e - \frac{1}{c^2} \frac{\partial ^2 p_e }{\partial t_{\mathbf{x}}^2}= 4 \pi G \sigma \rho_e,\,\frac{\partial }{\partial t_{\mathbf{x}}} \equiv \frac{1}{\beta (\mathbf{x},T)}\frac{\partial }{\partial T}.
\end{equation}
It is valid only in coordinates adapted to the preferred frame E. Moreover, $\sigma$ is defined as $\sigma \equiv T^{00}$. For this to be unambiguous, it is necessary to fix the time coordinate as $x^0 \equiv cT$ with $T$ the absolute time (see before Eq. (\ref{spacetimemetric})).

%%%%%%%%%%%%%%%%%%%%%%%%%%%%%%%%%%%%%%%%%%%%%%%%%%%%%%%%%%%%%%%%%%%
\section{Dynamics: Extension of Newton's 2nd law} \label{Dynamics}
%%%%%%%%%%%%%%%%%%%%%%%%%%%%%%%%%%%%%%%%%%%%%%%%%%%%%%%%%%%%%%%%%%%

Newton's second law: force = time-derivative of momentum, is the most general formulation of dynamics. It is more general than Lagrangian/Hamiltonian formulations (of which Einstein's assumption of a motion along space-time geodesics turns out to be a particular case). It is indeed a well-known fact in classical mechanics that Lagrangian systems are quite particular dynamical systems. Moreover, Newton's second law is based on a clear space/time separation, which corresponds to our intuitive concepts of space and time---whereas Einstein's assumption gives a physical status to space-time, and GR leads naturally to the possibility of time travels, \cite{Bonnor02} with their well-known paradoxes. Newton's second law has an obvious phenomenological flexibility, since we may alter the expression of the force. However, as regards the force of gravity, it should be equal to $m\mathbf{g}$ with $m$ the inertial mass and $\mathbf{g}$ the gravity acceleration, in order to save (at least for a {\it test particle}) the equality between inertial mass and passive gravitational mass. The latter equality, which is equivalent to the weak equivalence principle, is an accurately-established experimental fact, and is true for the investigated interpretation of gravity as Archimedes' thrust, as long as we stay in classical mechanics (see Subsect. \ref{Archimedes}). Now we account for SR, consistently with this interpretation, by saying that SR holds true {\it locally}---because an object moving through the macro-ether is Lorentz-contracted, as compared with an identical object staying at rest in the same place in the macro-ether made heterogeneous by gravity. Therefore, we must now take as the inertial mass the velocity-dependent mass $m(v)\equiv m(0)\gamma_v$, where the velocity $\mathbf{v}$ of the test particle, and its modulus $v$, are measured with clocks and rods of the momentarily-coincident observer bound to the macro-ether, thus with the physical space-time metric.  In the same way, the gradient operator entering Eq. (\ref{g_rho_e}) for $\mathbf{g}$ is now relative to the physical, curved space metric. Thus, the left-hand side of Newton's second law, that is the gravitational force, is unambiguously defined.\\

It is not trivial to define the {\it right-hand} side of Newton's second law, because, to this aim, we must define the time-derivative of a vector (the momentum $\mathbf{P} \equiv m(v)\mathbf{v}$) in the space endowed with the time-dependent Riemannian metric $\Mat{g}$. A unique definition has been derived for this time-derivative, from constraints which, I argue, must indeed be imposed. \cite{A16} This is in fact a mathematical result that applies independently of the theory of gravitation, although the definition had first been found in the framework of the investigated theory. \cite{A15} Thus, the r.h.s. of Newton's second law is uniquely defined for any theory of gravitation in a curved space-time. Einstein's geodesic motion is {\it characterized} by a particular form of the gravity acceleration vector $\mathbf{g}$, which depends, in the general case of a time-dependent metric, on the {\it velocity} of the particle as well as on its position. \cite{A16} Thus, geodesic motion takes place if and only if a particular form is assumed for $\mathbf{g}$. The ``if" part means that the four scalar equations involved in the geodesic equation are derived from the three ones in the extension of Newton's second law. \cite{A16} In the investigated theory, in which the assumed gravity acceleration (\ref{g_rho_e}) does {\it not} depend on the velocity of the test particle, geodesic motion is recovered for a constant gravitational field.\\

Dynamics being thus defined for a test particle, it is hence defined for a dust, which is a continuous medium made of non-interacting test particles. The dynamics of a test particle may be written in terms of its 4-acceleration vector, i.e. the absolute derivative of the 4-velocity $\mathbf{\mathsf{U}}=(U^\mu)$, \cite{A16} and in this form it is then immediately transcribed as an equation for the energy-momentum tensor $T^{\mu \nu } \equiv \rho^* U^\mu U^\nu$ of a dust ($\rho^*$ is the proper rest-mass density): independently of the assumed form for the space metric $\Mat{g}$ in the preferred frame, one gets~\cite{A20}
\begin{equation} \label{continuum}
T_{\mu;\nu}^{\nu} = b_{\mu},	
\ee
\be \label{b_mu}
b_0(\mathbf{T}) \equiv \frac{1}{2}\,g_{jk,0}\,T^{jk}, \quad b_i(\mathbf{T}) \equiv -\frac{1}{2}\,g_{ik,0}\,T^{0k}.	          
\ee
(Indices are raised and lowered with metric $\Mat{\gamma}$, and semicolon means covariant differentiation using the Christoffel connection associated with metric $\Mat{\gamma}$.) The universality of gravitation and the mass-energy equivalence mean exactly that this equation must hold true for a general continuous medium---be it a material or also a {\it nongravitational field}, such as the electromagnetic field. In particular, the modification of the Maxwell equations in a gravitational field is derived from this equation. \cite{B13} Equation (\ref{continuum}) with the definition (\ref{b_mu}) is valid in coordinates $(y^\mu)$ adapted to the preferred frame E and such that the time coordinate is $y^0 = \phi(T)$ with $T$ the absolute time (see before Eq. (\ref{spacetimemetric})). When combined with the equation for the scalar field [Eq. (\ref{champ1}) if the ``anisotropic" metric (\ref{spacemetric}) is assumed], Eq. (\ref{continuum}) implies a local conservation equation for the energy, which {\it substitutes} for the mass conservation of Newtonian theory: except for a dust, {\it mass is not exactly conserved}, although it is extremely close to be so in usual conditions.\cite{A20} This is an original prediction of the scalar theory.

%%%%%%%%%%%%%%%%%%%%%%%%%%%%%%%%%%%%%%%%%%%%%%%%%%%%%%%%%%%%%%%%%%%
\section{Observational agreement}\label{Experiment}
%%%%%%%%%%%%%%%%%%%%%%%%%%%%%%%%%%%%%%%%%%%%%%%%%%%%%%%%%%%%%%%%%%%

A survey of this topic has recently been written \cite{O2} for the scalar ether theory (with the ``anisotropic" space metric (\ref{spacemetric})). Although that theory is much simpler than GR, 
\footnote{ \
because: i) it is a scalar theory instead of a tensor one, ii) it has a fixed (and flat) background metric (whereas, in GR, even the space-time as a differentiable manifold cannot be considered given independently of the metric, \cite{Stachel} which leads to the necessity of adding gauge equations), and iii) there are no constraint equations to be imposed on the initial data (whereas, in GR, the time-time and time-space components of the Einstein equations actually impose four nonlinear constraints on the space metric and its time-derivative \cite{Rendall00}).
} 
it is more complex than Newtonian gravity (NG). Since the latter remains approximately valid for a weak gravitational field, one has to develop a ``post-Newtonian" (PN) approximation scheme, allowing to calculate the corrections to NG. This should be done in following the general principles of asymptotic analysis. Hence one should introduce a family of gravitational systems, depending on a field-strength parameter $\lambda$, and that family should be deduced from the data of the system of physical interest (e.g. the solar system), which system must correspond to a small value $\lambda_0$ of $\lambda$. 
\footnote{ \ 
This has been done for the scalar theory; it has been done in a very particular case and in an incomplete way for GR; see Ref. \cite{O1} and references therein.}
\\

The result of developing this ``asymptotic" PN scheme for the scalar theory is that the latter, in the ``anisotropic-metric" version, obtains the same predictions for the gravitational effects on light rays as the effects deduced in GR from the standard Schwarzschild metric. \cite{A19} (By the way, that metric is the unique solution of the theory in the static case with spherical symmetry. \cite{A19}) Moreover, the ``asymptotic" 1PN equations of motion of the mass centers of a weakly gravitating system have been derived and numerically implemented, and their coefficients have been least-squares-adjusted on a reference ephemeris: it seems that the ephemeris cannot be reproduced within $10''$ per century (see Ref. \cite{O2} and references therein). I argue that one should adjust the equations on direct observations (a hard work), for an ephemeris already represents a fitting of direct observations by equations derived from the reference theory (GR), using its standard (as opposed to ``asymptotic") PN scheme. \cite{O2} As to gravitational radiation, its analysis is based on an asymptotic post-Minkowskian approximation scheme, which leads to a ``quadrupole formula" very similar to that of harmonic GR, hence the data of binary pulsars should also be nicely fitted with the scalar theory. \cite{A34} Cosmology has also been investigated; in particular, the theory predicts that the cosmic expansion must be accelerated, and there is no singularity with infinite density. \cite{A28}

%%%%%%%%%%%%%%%%%%%%%%%%%%%%%%%%%%%%%%%%%%%%%%%%%%%%%%%%%%%%%%%%%%%
\section{Conclusion} \label{Conclusion} %%%%%%%%%%%%%%%%%%%%%%%%%%%%%%%%%%%%%%%%%%%%%%%%%%%%%%%%%%%%%%%%%%%

Euler's final comments on his interpretation of gravity as Archimedes' thrust in a fluid ``ether," translated at the end of Sect. \ref{Euler_pressure}, seem to remain relevant today. Newton's instantaneous attraction at a distance is just as impossible to ``physically understand" as it ever was, also for Newton himself. The necessity of recovering Newton's theory as a weak-field limit is a constraint on all theories (including GR) that aim at modifying Newton's theory to make it match with SR. By itself, this constraint does not bring any improvement to our physical understanding of gravity. However, it fits naturally with the physical concept of gravity being Archimedes' thrust due to the macroscopic pressure in a fluid {\it constitutive} ether (i.e., assumed to be the unique substance in the Universe, and of which the elementary particles would be mere local organizations). In this context, this constraint leads indeed to attribute non-Newtonian properties to gravity as a result of the compressibility of the fluid. In the present paper, I have shown that neither Euler's ether (which is external to the ``basic particles" that make matter), nor the macroscopic fields in the constitutive ether, can obey classical fluid mechanics. This reinforces the interpretation of the preferred reference frame of the theory, \cite{A8}--\cite{A20} according to which that frame would be defined by the average motion of the fluid ``micro-ether": the gravitational ether or ``macro-ether" does not obey mechanics, for its motion defines the very reference frame, in which mechanics is primarily written.\\

Since gravitation is intrinsically a result of the heterogeneity of ``space" (i.e., in the language of the theory, of the pressure or equivalently of the density in the macro-ether), whereas special relativity assumes that space is homogeneous, it follows that SR can apply only locally in the presence of gravitation. And since the Lorentz contraction and time-dilation in SR can be interpreted as resulting from a variation in the ``apparent" ether density, one is naturally led to postulate gravitational contraction and time-dilation. In this regard, the theory has been extended here, by allowing that the gravitational contraction can be isotropic, instead of occuring only in the direction of the gravity acceleration as previously assumed. This should solve the difficulties found \cite{A33,B23} with the weak equivalence principle. It has also been displayed the freedom which is left on the equation for the scalar gravitational field (the ``ether pressure"). What is imposed is the dynamics, which is defined by an extension of Newton's second law, and that dynamics together with the scalar field equation must imply an energy conservation. \cite{A15} How this can be done with the isotropic gravitational contraction, will be shown soon.

\bibliographystyle{amsplain}

\begin{thebibliography}{0}

{\small 
\bibitem{Newton1} I. Newton, {\it The Mathematical Principles of Natural Philosophy} (1st English edition, Benjamin Motte, London 1729), end of the ``Definitions" section.

\bibitem{Lunteren} F. van Lunteren, ``Nicolas Fatio de Duillier on the mechanical cause of universal gravitation," in Ref. \cite{Edwards}, pp. 41--59.

\bibitem{Edwards} M. R. Edwards (editor), {\it Pushing Gravity, New perspectives on Le Sage's theory of gravitation} (Apeiron, Montr\'eal 2002).

\bibitem{Euler1}
L. Euler, ``Recherches physiques sur la nature des moindres parties de la 
mati\`ere," in {\it Leonhardi Euleri Opera Omnia, Series Tertia, Pars Prima} (B. G. Teubner, Leipzig and Bern 1911), pp. 3--15 .­

\bibitem{Euler2}
L. Euler, ``Von der Schwere und den Kr\"aften so auf die himmlischen K\"orper 
wirken," in {\it Leonhardi Euleri Opera Omnia, Series Tertia, Pars Prima} (B. G. Teubner, Leipzig and Bern 1911), pp. 149--156.­

\bibitem{A8} 
M. Arminjon, ``A theory of gravity as a pressure force. I. Newtonian space and time," {\it Rev. Roum. Sci. Tech. -- M\'ec. Appl.} {\bf 38}, 3--24 (1993). 

\bibitem{A9} 
M. Arminjon, ``A theory of gravity as a pressure force. II. Lorentz contraction and 'relativistic' effects," {\it Rev. Roum. Sci. Tech. -- M\'ec. Appl.} {\bf 38}, 107--128 (1993). 

\bibitem{A15}
M. Arminjon, ``Energy and equations of motion in a tentative theory of gravity with a privileged reference frame," {\it Arch. Mech.} {\bf 48}, 25--52 (1996). 

\bibitem{A20}
M. Arminjon, ``On the possibility of matter creation/destruction in a variable gravitational field," {\it Analele Univ. Bucure\c{s}ti -- Fizica} {\bf 47}, 3--24 (1998). [physics/9911025]

\bibitem{Romani}
L. Romani, {\it Th\'eorie G\'en\'erale de l'Univers Physique (R\'eduction \`a la Cin\'ematique)} (Blanchard, Paris 1975 (Vol. 1) and 1976 (Vol. 2)).

\bibitem{PodlahaSjodin} 
M. F. Podlaha \& T. Sj\"odin, ``On universal fluids and de Broglie's waves," {\it Nuovo Cimento} {\bf 79B}, 85--92 (1984). 

\bibitem{L&L_fluides}
L. Landau \& L. Lifchitz, {\it M\'ecanique des Fluides} (Mir, Moscow 1971). [Original Russian edition: {\it Gidrodinamika} (Izd. Nauka, Moskva 1953).]

\bibitem{A28}
M. Arminjon, ``Accelerated expansion as predicted by an ether theory of gravitation," {\it Phys. Essays} {\bf 14}, 10--32 (2001). [gr-qc/9911057]

\bibitem{Will}
C. M. Will, {\it Theory and Experiment in Gravitational Physics} (2nd. edn., Cambridge University Press, Cambridge 1993).

\bibitem{Einstein1905} 
A. Einstein, ``Zur Elektrodynamik bewegter K\"orper," {\it Ann. der Phys.} (4) {\bf 17}, 891--921 (1905). (Received on June 30, 1905.) 

\bibitem{Logunov95} A. A. Logunov, {\it On the Articles by Henri Poincar\'e ``On the Dynamics of the Electron"} (English translation G. Pontecorvo, Publishing Department of the Joint Institute for Nuclear Research, Dubna 1995). (1st Russian edition 1984.) This monograph includes an English translation of Refs. \cite{Poincare1905,Poincare1906}.
 
\bibitem{Poincare1905} H. Poincar\'e, ``Sur la dynamique de l'\'electron," {\it C.-R. Acad. Sci. Paris} {\bf 140}, 1504--1508 (1905). (Presented at the session of June 5, 1905.)

\bibitem{Poincare1906} H. Poincar\'e, ``Sur la dynamique de l'\'electron," {\it Rendiconti Circ. Matemat. Palermo} {\bf 21}, 129--176 (1906). (Presented at the session of July 23, 1905.)

\bibitem{Lorentz} H. A. Lorentz, ``Electromagnetic phenomena in a system moving with any velocity smaller than that of light," {\it Proc. Acad. Sci. Amsterdam} {\bf 6}, 809--831 (1904). 

\bibitem{Prokhovnik67} S. J. Prokhovnik, {\it The Logic of Special Relativity} (Cambridge University Press, Cambridge 1967).

\bibitem{Prokhovnik85} S. J. Prokhovnik, {\it Light in Einstein's Universe} (Reidel, Dordrecht 1985).

\bibitem{Janossy} L. J\'anossy, {\it Theory of Relativity Based on Physical Reality} (Akad\'emiai Kiad\'o, Budapest 1971).

\bibitem{Pierseaux} Y. Pierseaux, {\it La ``Structure Fine" de la Relativit\'e Restreinte} (L'Harmattan, Paris 1999). 

\bibitem{Brandes01} J. Brandes, {\it Die relativistischen Paradoxien und Thesen zu Raum und Zeit} (3rd edition, Verlag relativistischer Interpretationen, Karlsbad 2001)(1st edition 1994).

\bibitem{A18} 
M. Arminjon, ``Scalar theory of gravity as a pressure force," {\it Rev. Roum. Sci. Tech. -- M\'ec. Appl.} {\bf 42}, 27--57 (1997). 

\bibitem{MansouriSexl}
R. Mansouri \& R. Sexl, ``A test theory of special relativity. I. Simultaneity and clock synchronization. {\it Gen. Rel. Gravit.} {\bf 8}, 497--513 (1977); ``A test theory of special relativity. II. First-order tests," {\it Gen. Rel. Gravit.} {\bf 8}, 515--524  (1977); ``A test theory of special relativity. III. Second-order tests," {\it Gen. Rel. Gravit.} {\bf 8},  809--814 (1977).

\bibitem{SelleriFP}
F. Selleri, ``Noninvariant one-way velocity of light," {\it Found. Phys.} {\bf 26}, 641--664 (1996); ``Noninvariant one-way velocity of light and particle collisions," {\it Found. Phys. Lett.} {\bf 9}, 43--60 (1996); ``On the Fizeau experiment," {\it Found. Phys. Lett.} {\bf 16},  71--82  (2003). F. Selleri and R. Manaresi, ``The international atomic time and the velocity of light," {\it Found. Phys. Lett.} {\bf 17},  65--79 (2004).

\bibitem{Selleri03}
F. Selleri, {\it Lezioni di Relativit\'a da Einstein all'etere di Lorentz}, Progedit, Bari (2003). English translation in preparation.

\bibitem{A19}
M. Arminjon, ``Post-Newtonian approximation of a scalar theory of gravitation and application to light rays," {\it Rev. Roum. Sci. Tech. -- M\'ec. Appl.} {\bf
43}, 135--153 (1998). [gr-qc/9912041]

\bibitem{A33}
M. Arminjon, ``Equations of motion of the mass centers in a scalar theory of gravitation: the point particle limit," submitted for publication. [gr-qc/0301031]

\bibitem{B23}
  M. Arminjon, ``Point-particle limit in a scalar theory of gravitation and the weak equivalence principle," in {\it Gravitational Waves and Experimental Gravity}, Proc. 38th Rencontres de Moriond (J. Dumarchez \& J. Tr\^an Thanh V\^an, eds., The Gioi, Hanoi 2004), pp. 377--382. [gr-qc/0306025]

\bibitem{O2}
M. Arminjon, ``Ether theory of gravitation: why and how?" in {\it Relativistic World Ether} (M. C. Duffy \& L. Kostro, eds.), book to be published. [gr-qc/0401021]

\bibitem{O1}
M. Arminjon, ``Testing a theory of gravity in celestial mechanics: a new method and its application to a new scalar  theory,"  in {\it Recent Research Developments in Astronomy \& Astrophysics} (S. G. Pandalai, ed., Research Sign Post, Trivandrum), Vol. 1 (2003), pp. 859-879. [gr-qc/0305078]

\bibitem{Bonnor02} W. B. Bonnor, ``Closed time-like curves," Communication at the {\it Second British Gravity Meeting}, Queen Mary University of London, June 10/11, 2002; gr-qc/0211051.

\bibitem{A16}
M. Arminjon, ``On the extension of Newton's second law to theories of gravitation in curved space-time," {\it Arch. Mech.} {\bf 48}, 551--576 (1996).

\bibitem{B13}
M. Arminjon, ``Gravitation as a pressure force: a scalar ether theory," in {\it Physical Interpretations of Relativity Theory V, Supplementary Papers}, edited by M.C. Duffy
(Univ. of Sunderland/ Brit. Soc. Philos. Sci., Sunderland 1998), pp.
1--27. 

\bibitem{Stachel}
J. Stachel, ``Einstein's search for general covariance," in {\it Einstein and the History of General Relativity}, edited by D. Howard \& J. Stachel (Birkha\"user, Boston-Basel-Berlin 1989), pp. 63-100.

\bibitem{Rendall00}
A. D. Rendall, ``Local and global existence theorems for the Einstein equations," {\it Living Reviews in Relativity} {\bf 3} (2000),
1, article on the web at \\
\verb+http://www.livingreviews.org/Articles/Volume3/2000-1rendall+.

\bibitem{A34}
M. Arminjon, ``Gravitational effects on light rays and binary pulsar energy loss in a scalar theory of gravity,"  {\it Theor. Math. Phys.} {\bf 140} (1), 1011-1027 (2004) [{\it Teor. Mat. Fiz.} {\bf 140}  (1), 139-159 (2004)]. [gr-qc/0310062]


}
%}
\end{thebibliography}
%\bibliography
%{

\end{document}